\documentclass[%
 aip,%
 amsmath,amssymb,
 reprint,%
]{revtex4-1}

\usepackage{graphicx}
\usepackage{dcolumn}
\usepackage{bm}
\usepackage{xcolor}

\begin{document}


\title[]{Note: On the dielectric constant of nanoconfined water}

\author{Chao Zhang}
\email{chao.zhang@kemi.uu.se}
\affiliation{Department of Chemistry-\AA ngstr\"om Laboratory, Uppsala
  University, L\"agerhyddsv\"agen 1, 75121, Uppsala, Sweden}


\maketitle

%


Investigations of dielectric properties of liquid water in
nanoconfinement are highly relevant for the energy storage in
electrochemical systems, 
mineral-fluid interactions in geochemistry and
microfluid based devices in biomedical analysis~\cite{Bjorneholm:2016jb}. It has been
reported that polarization shows a strong anisotropy at water
interfaces~\cite{Zhang:2013wl, DeLuca:2016jp} and
the dielectric constant of nanoconfined water is surprisingly
low ($\epsilon_\perp \sim 10$)~\cite{Itoh:2015dg}. Here, using a simple capacitor model, we show that the low dielectric constant of
nanoconfined water can be largely accounted by the so-called
dielectric dead-layer
effect known for ferroelectric nanocapacitors~\cite{Stengel:2006kv}. 

Before talking about the effect of nanoconfinement, one needs to realize
that the first effect of having an interface corresponds to a switch in the
electric boundary condition. From classical electrodynamics, we know
that the electric field $E_z$ is discontinuous at a dielectric interface, that is the
reason why it is convenient to use the electric displacement $D$ as
the fundamental variable instead~\cite{Matyushov:2014fb}. In the latter case,
the polarization of dielectrics $P_z$ is added into the electric
field. This makes $D$ continuous  in the direction perpendicular to an
interface and leads to its definition $D=E_z+4\pi P_z$.

When the electric boundary condition is switched from constant
electric field $E$ to constant electric displacement $D$, the
dielectric response will be different
accordingly~\cite{Bonthuis:2012ba, Anonymous:2016ij, Zhang2016a,Zhang2016c}. 

\begin{eqnarray}
P_{\parallel}=\chi_{\parallel} E&,& \epsilon_{\parallel}=1+4\pi\chi_{\parallel} \label{chi_para}\\
P_{\perp}=\chi_{\perp} D&,& \epsilon_{\perp}=1/(1-4\pi\chi_{\perp}) \label{chi_perp}
\end{eqnarray}

The difference in $\chi$ due to the electric
boundary condition leads to differences in the fluctuation of
polarization at zero field and in the corresponding
relaxation time. This phenomenon is not limited to water in nanoconfinement where
switching of electric boundary condition is enforced by introducing
explicit interfaces~\cite{Zhang:2013wl, DeLuca:2016jp} but can also be
realized in bulk liquid water by turning on the constant electric displacement
simulation~\cite{Zhang2016a,Zhang2016c}. For bulk liquid water, $\epsilon_{\perp}=\epsilon_{\parallel}$ even
though $\chi_{\perp}$ is radically different from $\chi_{\parallel}$. 

Now let us go back to the original question: What accounts for the low dielectric constant
$\epsilon_{\perp}$ of water slab in nanoconfinement~\cite{Itoh:2015dg}. Is water in
nanoconfinement completely different from that in the bulk ? 

\begin{figure}
\includegraphics[width=0.95\columnwidth]{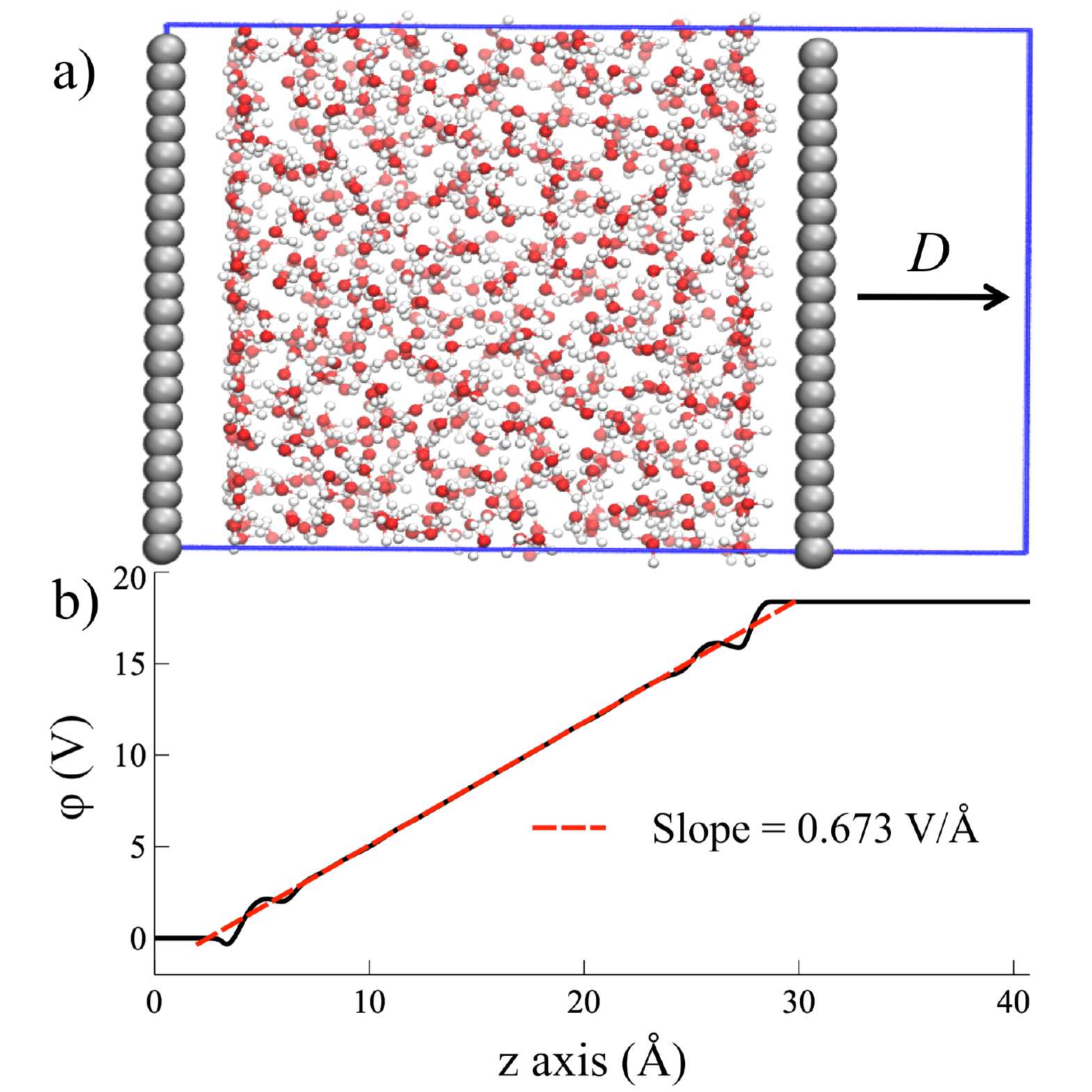}
\caption{\label{fig1} a) A snapshot of MD simulations of water slab
  confined between rigid walls under constant electric displacement $D=0.6835$V/\AA~. The seperation distance between walls
  $L_\text{w}$ is 30.77\AA~in this case.; b) The corresponding electrostatic potential
profile $\varphi(z)$ generated from the charge
density. The slope gives the negative of the deploarization field $4\pi
P_{\perp}$ in the bulk water region.}
\end{figure}

To answer this question, we applied the constant electric displacement
simulation with $D=0.6835$V/\AA~to a water slab confined between two
hydrophobic walls at ambient conditions (Fig.~\ref{fig1}a). Interactions between water molecules is described
by simle point charge/extended (SPC/E) model~\cite{Berendsen:1987uu}
and the rigid hydrophobic walls
are composed of atoms on a dense cubic lattice. All molecular dynamics
(MD) simulations were performed
with GROMACS 4 package~\cite{Hess2008} and technical settings are the same as
described in the previous work~\cite{Zhang2016a}. 

From the corresponding electrostatic potential profile $\varphi(z)$,
we can extract the deploarization field $-4\pi P_{\perp}$ from the
slope in the middle region of water slab (Fig.~\ref{fig1}b). Inserting
this value into Eq.~\ref{chi_perp} and knowing $D=0.6835$V/\AA~as
the control variable, one gets $\epsilon_{\perp,\text{bulk}}=65$. This
number is indeed quite close to that of the bulk
liquid water at the same magnitude of $D$, see Ref.~\cite{Zhang2016a}.  In other words, water
slab of about 30~\AA~thick can already recover the bulk dielectric
response. Then, the question is why the reported dielectric constant
$\epsilon_{\perp}$ can be as low as a single digit number~\cite{Itoh:2015dg}?

One needs to realize that $\epsilon_{\perp}$  is the overall dielectric
constant which includes both surface contribution and bulk
contribution. Because the simulation system is under the constant
electric displacement condition, therefore surface region and bulk
region can be regarded as capacitors connected in series. This was
already pointed out in the study of ferroelectric nanocapacitor~\cite{Stengel:2006kv}.

\begin{equation}
\label{capacitor}
\frac{1}{C_{\perp}}=\frac{1}{C_{\text{surf}}}+\frac{1}{C_{\text{bulk}}}
\end{equation} 

where $C_{\perp}=\epsilon_{\perp}/L_\text{w}$,
$C_\text{surf}=\epsilon_{\perp, \text{surf}}/L_\text{surf}$ and
$C_\text{bulk}=\epsilon_{\perp,
  \text{bulk}}/(L_\text{w}-L_\text{surf})$. $L_\text{w}$ is the seperation distance
between walls and $L_\text{surf}$ is the total width of two interfaces.

Because of this sum of inverses, the region which has a smaller dielectric
constant will dominate. From
Fig.~\ref{fig1}, one can clear see there are two vacuum gaps between walls and
confined water slab. Therefore, our simple capacitor
model will just approximate the confined water as vacuum gaps ($\epsilon_{\perp, \text{surf}}=1$)
plus bulk water ($\epsilon_{\perp,\text{bulk}}=65$ at $D=0.6835$V/\AA). 

\begin{figure} [h]
\includegraphics[width=0.95\columnwidth]{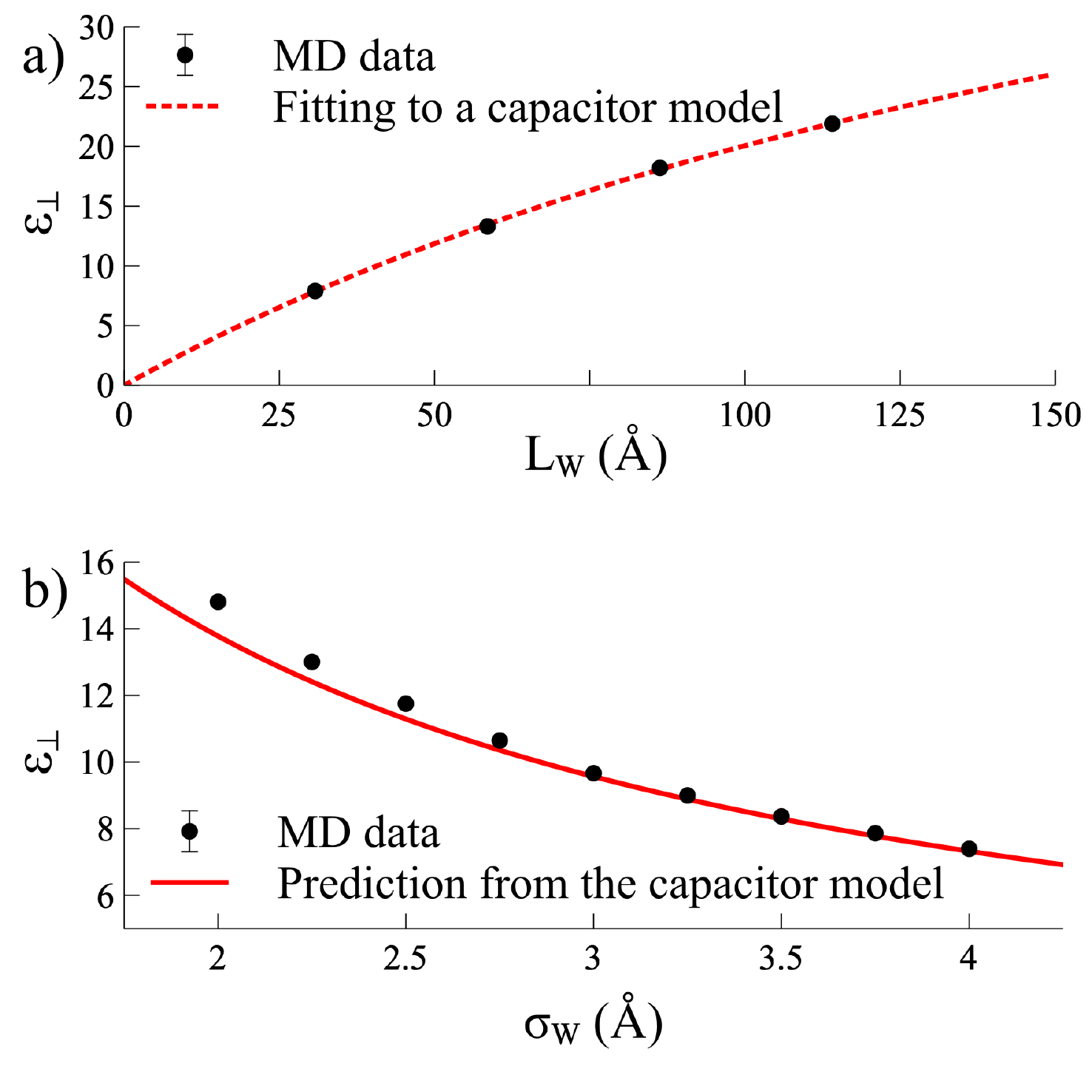}
\caption{\label{fig2} a) The dielectric constant of nanoconfined water
  $\epsilon_{\perp}$ as a function of the seperation distance $L_\text{w}$. Lennard-Jones potential parameter $\sigma_\text{w}=3.75$\AA~for wall atoms. The capacitor model
  corresponds to Eq.~\ref{epsilon_tot}. The
  only free parameter is $\sigma_\text{corr}$. The error of
  $\epsilon_{\perp}$ calculated from MD simulations was estimated from the block average and negligibly small.; b) The predicated dielectric constant of nanoconfined water
  $\epsilon_{\perp}$ as a function of Lennard-Jones potential parameter
  of wall atoms  $\sigma_\text{w}$. $L_\text{w}=30.77$\AA, 
  $\epsilon_{\perp,\text{bulk}}$ and $\sigma_\text{w}$  are all model
  parameters.}
\end{figure}

Base on these considerations and Eq.~\ref{capacitor}, $\epsilon_{\perp}$ can be rewritten as:

\begin{equation}
\label{epsilon_tot}
\epsilon_{\perp}=\frac{L_\text{w}}{L_\text{surf}+(L_\text{w}-L_\text{surf})/\epsilon_{\perp, \text{bulk}}}
\end{equation}

Here, the only unknown parameter is $L_\text{surf}$, which is the width of
vacuum gaps in this capacitor model. Because  $L_\text{surf}$ depends
on the van der Waals radius of wall atoms and interfacial water
molecules, therefore, we approximate $L_\text{surf}$ as
$\sigma_\text{w}+\sigma_\text{corr}$.  $\sigma_\text{w}$ is the 
interatomic distance when the underlying Lennard-Jones potential becomes zero and
it roughly doubles the van der Waals radius of the
corresponding wall atom. Because water in the nanoconfined geometry face
two walls, therefore we consider $\sigma_\text{w}$ as a first
approximation of $L_\text{surf}$. The remaining term 
$\sigma_\text{corr}$ is a correction factor for the mixing effect, thus it should be small and can be obtained by fitting MD data.

Results of $\epsilon_{\perp}$ as a function of $L_\text{w}$ are shown in
Fig.~\ref{fig2}a. Fitting MD data with Eq.~\ref{epsilon_tot} gives
$\sigma_\text{corr}$ as 0.22\AA, which turns out to be small as supposed. Using this model, one can make a prediction
regarding the relationship between the Lennard-Jones parameter
$\sigma_\text{w}$ of wall atoms and the dielectric constant of
nanoconfined water $\epsilon_{\perp}$. The agreement with MD
simulations is encouragingly good (Fig.~\ref{fig2}b). We conclude that this simple
capacitor model successfully captures the main physical reason behind
the low dielectric constant  $\epsilon_{\perp}$ of water in
nanoconfinement. Nevertheless, one should be aware that the dielectric
constant of the interface $\epsilon_{\perp, \text{surf}}$ is (drastically) approximated to be
1 in Eq.~\ref{capacitor} and future works should include the effect of the interfacial water (Fig.~\ref{fig1}b).

%


\end{document}